\documentclass[useAMS,usenatbib,usegraphicx,usedcolumn]{mn2e}
\usepackage{amssymb,amsmath}
\usepackage{times}
\title[Inverse-Compton ghosts and giant radio sources]{Inverse-Compton ghosts and double-lobed radio sources in the X-ray sky}
\author[P. Mocz, A.C. Fabian \& K.M. Blundell]{P. Mocz$^{1,2}$\thanks{E-mail: pmocz@fas.harvard.edu (PM); acf@ast.cam.ac.uk (ACF); kmb@astro.ox.ac.uk (KMB)},
 A.C. Fabian$^{2}$\footnotemark[1] and Katherine M. Blundell$^{3}$\footnotemark[1]\\
$^{1}$Harvard University, Cambridge, MA 02138, USA\\
$^{2}$Institute of Astronomy, Madingley Road, Cambridge CB3 0HA\\
$^{3}$Astrophysics, University of Oxford, Keble Road, Oxford OX1 3RH}
\begin{document}

\date{MNRAS accepted 13 December 2010. Subm. 12 August 2010.}

\pagerange{\pageref{firstpage}--\pageref{lastpage}} \pubyear{2010}

\maketitle

\label{firstpage}

\begin{abstract}
In this study we predict the total distributions of powerful (FR II) active double-lobed radio galaxies and ghost sources, and their observable distribution in the X-ray sky. We develop an analytic model for the evolution of the lobe emission at radio and X-ray energies. During jet activity, a double radio source emits synchrotron radiation in the radio and X-ray emission due to inverse-Compton (IC) upscattering by $\gamma\sim10^3$ electrons of the cosmic microwave background. After the jets switch off, the radio luminosity (due to higher $\gamma$ electrons) falls faster than the X-ray luminosity and for some time the source appears as an IC ghost of a radio galaxy before becoming completely undetectable in the X-ray. With our model, for one set of typical parameters, we predict radio lobes occupy a volume fraction of the universe of $0.01$, $0.03$, $0.3$ at $z=2$ (during the quasar era) of the filamentary structures in which they are situated, for typical jet lifetimes $5\times 10^7$~yr, $10^8$~yr, $5\times 10^8$~yr; however since the inferred abundance of sources depends on how quickly they fall below the radio flux limit the volume filling factor is found to be a strong function of radio galaxy properties such as energy index and minimum $\gamma$ factor of injected particles, the latter not well constrained by observations. We test the predicted number density of sources against the {\it Chandra} X-ray Deep Field North survey and also find the contribution to the unresolved cosmic X-ray background by the lobes of radio galaxies. $10$--$30$~per~cent of observable double-lobed structures in the X-ray are predicted to be IC ghosts. The derived X-ray luminosity function of our synthetic population shows that double-lobed sources have higher space densities than X-ray clusters at redshifts $z\geq 2$ and X-ray luminosities above $10^{44}$~erg~s$^{-1}$.
\end{abstract}

\begin{keywords}
galaxies: active -- galaxies: evolution -- galaxies: jets -- radio continuum: galaxies -- X-rays: galaxies.
\end{keywords}

\section{Introduction}\label{Introduction}

We investigate how much of intergalactic space is filled with current and old radio lobes. 
Most galaxies seem to have a supermassive black hole at their centres \citep{1998AJ....115.2285M}, which can serve as a central engine for jet ejection.
It is possible that most massive galaxies have at least one outburst of jet activity leading to a giant radio source in their lifetime (lasting $\sim10^8~{\rm yr}$), probably between redshifts $1.5$ to $3$ during the quasar era \citep{2001ApJ...560L.115G}. 
The radio galaxies will have extended emission in the radio as well as extended X-ray emission in the keV energies due to inverse-Compton (IC) upscattering by $\gamma\sim10^3$ electrons of photons that comprise the cosmic microwave background (CMB). The IC and synchrotron losses will downshift the higher energy electrons ($\gamma\sim10^4$) responsible for GHz synchrotron radiation more quickly than the lower-energy electrons that give rise to the X-ray emission. Thus for some period of time after the jet is switched off the source will appear as an inverse-Compton ghost of a radio source before becoming completely radio and X-ray dark. 
The extended X-ray source HDF 130 ($z=1.99$) has been interpreted to be such a radio galaxy with its jets turned off and only showing a double-lobed structure in the X-ray \citep{2009MNRAS.395L..67F}, and the models developed in this paper are applied to that object in a companion paper. The CMB energy density is proportional to $(1+z)^4$, cancelling the dimming due to distance, and so extended X-ray emission may be observable at both low and high redshifts. Previous work on the extended X-ray emission of radio galaxies range from as early as \cite{1969Natur.221..924F}, which described the cosmic X-ray background (CXB) being attributed to a Compton-blackbody process acting in double radio sources, to \cite{1989MNRAS.239..173G}, which is an analytic treatment of the time evolution of the radio output of expanding radio lobes incorporating IC losses, to \cite{2004MNRAS.353..523C}, which considered the relative abundance of double radio galaxies to X-ray clusters, and recently to the work by \cite{2010MNRAS.tmp.1004N} which developed a simple model for the evolution of the X-ray luminosities of double radio sources. 

We create an analytic model accounting for injection of relativistic particles into the lobes of radio galaxies and adiabatic, synchrotron and IC energy losses to describe the evolution of emission in the radio and the X-ray bands of a double-lobed radio galaxy which turns off after a typical jet lifetime. We can describe the total population of active (i.e. jets switched on) and no longer active (jets off) sources from an empirically-inferred birth function and jet energy distribution. Of interest are the volume filling factor of the universe or filaments by radio lobes, their observable number density in X-rays, their contribution to the unresolved CXB and the X-ray luminosity function (XLF) of these extended X-ray sources. We explore and find that individual radio galaxy properties such as the index and minimum energy of the injection spectrum, and the ambient density, can significantly affect the time at which radio sources fall below a given flux limit and consequently have a significant impact on the actual observed distribution of sources.

We consider the distribution and evolution of powerful Fanaroff and Riley Class II (FR II) \citep{1974MNRAS.167P..31F} sources to understand what fraction of filaments may be filled with radio lobes. We consider a cosmology with $H_0=70~{\rm km}~{\rm s}^{-1}~{\rm Mpc}^{-1}$, $\Omega_{\rm M}=0.27$, $\Omega_\Lambda=0.73$.

\section{The Model}\label{sec:model}

In order to study the evolution of the lobes of FR II radio galaxies, we create an analytic model where the lobes are fed with synchrotron emitting electrons from a compact hotspot which undergo adiabatic expansion and synchrotron and IC losses. We follow the formalisms found in \cite{1997MNRAS.286..215K}, \cite{1997MNRAS.292..723K}, \cite{1999AJ....117..677B} and \cite{2010MNRAS.tmp.1004N}. We are interested in studying the evolution of X-ray and radio emission of these sources to beyond the jet lifetime $t_{\rm j}$.

We pick one set of typical model parameters (case [A]) describing the ambient environment and injection spectrum based on  parameters used in previous works by \cite{1997MNRAS.286..215K}, \cite{1997MNRAS.292..723K}, \cite{1999AJ....117..677B} and \cite{2010MNRAS.tmp.1004N}. 
However, we later show the importances individual parameters may have in altering the inferred distribution of sources (\S~\ref{sec:params}).
Case [A] consists of parameters inferred from observations and similar to parameters used in previous models for radio sources, and so it serves as a normative case.
However, not all parameters are well constrained by observations, such as $\gamma_{\rm min}$ (see discussion on this in \S~\ref{sec:modpar}), and so some of the other cases from [B]-[L] may also be quite plausible models for a typical radio lobe.

\subsection{Synchrotron radiation basics}\label{sec:basics}

The synchrotron luminosity $\delta P_\nu$ in units of ${\rm W}~{\rm Hz}^{-1}~{\rm sr}^{-1}$, averaged over pitch angle, of a volume element $\delta V$ with magnetic energy density $u_{\rm B}$ is given by:
\begin{equation}
\delta P_\nu =\frac{1}{6\pi}\sigma_{\rm T} c u_{\rm B} \frac{\gamma^3}{\nu}n(\gamma)\delta V
\label{eq:synchrotron}
\end{equation}
where $\gamma$ is the Lorentz factor associated with an electron radiating at frequency $\nu$, $n(\gamma)d\gamma$ is the number density of electrons with Lorentz factors between $\gamma$ and $\gamma+d\gamma$, $\sigma_{\rm T}$ is the Thomson scattering cross section and $c$ is the speed of light. We assume that electrons only emit at their peak frequency given by $\nu=\gamma^2\nu_{\rm L}$, where $\nu_{\rm L}$ is the Larmor frequency.

The energy loss equation describes the time evolution of the Lorentz factor:
\begin{equation}
\frac{d\gamma}{dt}=-\gamma\frac{1}{3}\frac{1}{V_{\rm l}}\frac{dV_{\rm l}}{dt}-\frac{4}{3}\frac{\sigma_{\rm T}}{m_{\rm e} c}\gamma^2 (u_{\rm B}+u_{\rm c})
\label{eq:loss}
\end{equation}
where the first term is the energy loss due to the adiabatic expansion of the lobe of volume $V_{\rm l}$ and the second term describes the synchrotron and IC losses. Here $m_{\rm e}$ is the mass of an electron and $u_{\rm c}=a(2.7~{\rm K}~(1+z))^4$ is the CMB photon energy density at the redshift of the source. 
$a=7.565\times10^{-16}~{\rm J}~{\rm K}^{-4}~{\rm m}^{-3}$ is the radiation constant.

Supposing we know the distribution of electrons $n(\gamma_{\rm i},t_{\rm i})$ as initially injected into the lobe at time $t_{\rm i}$ and assuming the electrons are uniformly distributed over volume, we can find the distribution of those electrons at a later time, $n(\gamma,t)$, from:
\begin{equation}
n(\gamma,t)d\gamma=n(\gamma_{\rm i},t_{\rm i})\frac{V_{\rm l}(t_{\rm i})}{V_{\rm l}(t)}\frac{d\gamma_{\rm i}}{d\gamma}d\gamma
\label{eq:gammaEvo}
\end{equation}
where determining $\gamma_{\rm i}$ and $d\gamma_{\rm i} / d\gamma$ is described in \S~\ref{sec:RadioL}.

\subsection{Dynamics of the radio lobes}\label{sec:dynamics}

The density of the environment surrounding a radio galaxy will affect both the dynamics and energy evolution of the jets and lobes. We assume a density profile around the AGN described by
\begin{equation}
\rho(r)=\rho_0(r/a_0)^{-\beta}
\label{eq:environment}
\end{equation}
where for case [A] we take $\beta=1.5$, $a_0=10~{\rm kpc}$ and $\rho_0=1.67\times 10^{-23}~{\rm kg}~{\rm m}^{-3}$ \citep{1999AJ....117..677B}, inferred from observations, but will also consider steeper and less dense profiles (\S~\ref{sec:params}). It may be the case that the radio-source environments change with redshift, but we will assume independence from $z$ as the evolution of radio source environments is not yet fully understood.

We assume in our model a double jet system with origin at the central engine of an AGN, each jet transporting a total power $Q_{\rm j}$, assumed to be constant while the jet is on. The jet will have a lifetime of $t_{\rm j}$, after which the jet activity stops and no further particles are injected into the lobe. During a time $t<t_{\rm j}$, we use the source expansion based on the characteristic length scale $(t^3Q_{\rm j}\rho_0^{-1}a_0^{-\beta})^{1/(5-\beta)}$ described in \cite{1991MNRAS.250..581F}:
\begin{equation}
L_{\rm j}(t)=c_1\left(\frac{t^3Q_{\rm j}}{\rho_0a_0^\beta}\right)^{1/(5-\beta)}
\label{eq:Lj}
\end{equation}
where $L_{\rm j}(t)$ is the length of a jet and $c_1$ is a dimensionless constant taken to be $1.8$ as in \cite{1999AJ....117..677B}. The total length of the source is $2L_{\rm j}$.

We make the assumptions of \cite{1999AJ....117..677B} that the jet remains relativistic from the central engine to the jet shock (i.e. hotspot), that the jet has lower density than the surrounding environment, that at the outermost edge of the source at $L_{\rm j}(t)$ is the hotspot just beyond the shock structure where bulk kinetic energy from the jet is randomized feeding the larger head (outermost bright emission region) of the lobe. The head pressure is responsible for the rate at which the source grows. The hotspot pressure is responsible for adiabatic expansion losses out of the hotspot (see \cite{1999AJ....117..677B}).

While the jet is on, the pressure in the head can be found from the jump conditions of a strong shock, as discussed in \cite{1997MNRAS.292..723K}; we have:
\begin{equation}
p_{\rm head}(t)=\frac{18 \rho_0 c_1^{2-\beta}}{(\Gamma_{\rm x}+1)(5-\beta)^2}a_0^2\left(\frac{Q_{\rm j}}{a_0^5\rho_0}\right)^{\frac{2-\beta}{5-\beta}}t^{-\frac{4+\beta}{5-\beta}}
\label{eq:phead}
\end{equation}
where $\Gamma_{\rm x}=5/3$ is the adiabatic index of the surrounding IGM.

The lobe is treated as consisting of small volume elements $\delta V$, each with varying fluid properties. A fluid element is injected at some time $t_{\rm i}$ from the hotspot over a time interval $\delta t_{\rm i}$. The element $\delta V(t_{\rm i})$ is related to $\delta t_{\rm i}$ by:
\begin{equation}
\delta V(t_{\rm i})=\frac{(\Gamma_{\rm l}-1)Q_{\rm j}}{p_{\rm l}(t_{\rm i})}\left(\frac{Q_{\rm j}}{c A_{\rm hs} p_{\rm l}(t_{\rm i})}\right)^{\frac{1-\Gamma_{\rm l}}{\Gamma_{\rm l}}}\delta t_{\rm i}
\label{eq:dVi}
\end{equation}
using thermodynamic relations assuming adiabatic expansion of the volume element over the time interval (equation (19) in \cite{1999AJ....117..677B}). $A_{\rm hs}$ is the area of the hotspot (the hotspot is assumed to have a fixed radius of $2.5~{\rm kpc}$), $\Gamma_{\rm l}=4/3$ is the adiabatic index of the lobe and $p_{\rm l}$ is the pressure in the lobe, which we equate with the pressure in the head given by equation~(\ref{eq:phead}) divided by a factor of $6$, adopted from \cite{1999AJ....117..677B}, to allow for the pressure gradient along the lobe. Then, assuming further adiabatic expansion, the volume element changes as:
\begin{equation}
\delta V(t)=\frac{(\Gamma_{\rm l}-1)Q_{\rm j}}{p_{\rm l}(t_{\rm i})}\left(\frac{Q_{\rm j}}{c A_{\rm hs} p_{\rm l}(t_{\rm i})}\right)^{\frac{1-\Gamma_{\rm l}}{\Gamma_{\rm l}}}\left(\frac{p_{\rm l}(t_{\rm i})}{p_{\rm l}(t)}\right)^{\frac{1}{\Gamma_{\rm l}}}\delta t_{\rm i}.
\label{eq:dV}
\end{equation}

The volume of each lobe at time $t$ can be found by integrating equation~(\ref{eq:dV}). 
That is, for $t\leq t_{\rm j}$:
\begin{equation}
V_{\rm l}(t)=\displaystyle\int_0^{t}\frac{(\Gamma_{\rm l}-1)Q_{\rm j}}{p_{\rm l}(t_{\rm i})}\left(\frac{Q_{\rm j}}{c A_{\rm hs} p_{\rm l}(t_{\rm i})}\right)^{\frac{1-\Gamma_{\rm l}}{\Gamma_{\rm l}}}\left(\frac{p_{\rm l}(t_{\rm i})}{p_{\rm l}(t)}\right)^{\frac{1}{\Gamma_{\rm l}}}\,dt_{\rm i}.
\label{eq:V}
\end{equation}
For a general time $t$, we want to set the integration limit to $\min [t,t_{\rm j}]$, as nothing is injected into the lobes once the jet has shut down.

We call the axial ratio of the lobe $R=R(t)$ (nucleus-hotspot distance divided by full width of the lobe), so that the volume of a lobe is:
\begin{equation}
V_{\rm l}(t)=\frac{\pi}{4R(t)^2}L_{\rm j}(t)^3
\label{eq:volume}
\end{equation}
 (the lobe is assumed to have a cylindrical shape). For $t\leq t_{\rm j}$, we know how $V_{\rm l}$ and $L_{\rm j}$ evolve and thus can determine how $R$ grows with expansion.

Now we will describe the evolution of the lobe after the jet has turned off. The expansion of the lobe for $t>t_{\rm j}$ is governed by 
\begin{equation}
\dot{L_{\rm j}}=\sqrt{\frac{p_{\rm l}}{\rho_{\rm a}}}
\label{eq:evo2}
\end{equation}
where $\rho_{\rm a}$ is the ambient density at $L_{\rm j}(t)$, which we know from equation~(\ref{eq:environment}).

It is not reasonable to assume that the source continues to expand according to equation~(\ref{eq:Lj}) once the jet activity has discontinued since at this point $Q_{\rm j}=0$  and we do not have a characteristic length. To proceed, we assume that the axial ratio for $t>t_{\rm j}$ will be given by its value at $R(t_{\rm j})$, that is, the axial ratio at the time when the jet turned off. 
This is a reasonable assumption if the axial ratio is small enough and the distance from the galaxy large enough so that the external pressure at the heads and sides of the jet are similar, even in an ambient environment with density described by a power-law decline. The assumption is useful for obtaining an analytic solution for the problem, and has been made for the entirety of the evolution of sources in other models such as those of \cite{1997MNRAS.286..215K}, \cite{1997MNRAS.292..723K}, and \cite{2010MNRAS.tmp.1004N}.
Noticing from equation~(\ref{eq:V}) that for $t>t_{\rm j}$ we have 
\begin{equation}
V_{\rm l}(t)=V_{\rm l}(t_{\rm j})\left(\frac{p_{\rm l}(t_{\rm j})}{p_{\rm l}(t)}\right)^{\frac{1}{\Gamma_{\rm l}}},
\label{eq:volevo}
\end{equation}
 and using equation~(\ref{eq:volume}), we find after substitution from equation~(\ref{eq:evo2}) that for $t>t_{\rm j}$,
\begin{equation}
\dot{L_{\rm j}}=k^{\frac{1}{2}}(L_{\rm j}^{-3\Gamma_{\rm l}/2+\beta/2})
\label{eq:diffeq}
\end{equation}
where
\begin{equation}
k\equiv\frac{p_{\rm l}(t_{\rm j})}{a_0^\beta \rho_0}\left(V_{\rm l}(t_{\rm j}) \frac{4R(t)^2}{\pi}\right)^{\Gamma_{\rm l}}.
\label{eq:k}
\end{equation}
With the assumption that $R(t)=R(t_{\rm j})$ for post-jet conditions, we find an analytic solution for the evolution of $L_{\rm j}$ at $t>t_{\rm j}$ by solving the differential equation~(\ref{eq:diffeq}) with appropriate boundary condition at time $t_{\rm j}$:
\begin{equation}
L_{\rm j}(t)=\left((3\Gamma_{\rm l}/2+1-\beta/2)(\sqrt{k}(t-t_{\rm j})+C) \right)^{2/(3\Gamma_{\rm l}+2-\beta)}
\label{eq:post L}
\end{equation}
with
\begin{equation}
C\equiv\frac{L_{\rm j}(t_{\rm j})^{3\Gamma_{\rm l}/2+1-\beta/2}}{3\Gamma_{\rm l}/2+1-\beta/2}.
\label{eq:C}
\end{equation}

From $L_{\rm j}(t)$ we can determine $V_{\rm l}(t)$ and consequently $p_{\rm l}(t)$ from equation~(\ref{eq:volevo}), for $t>t_{\rm j}$.

\subsection{Radio luminosity of the lobe}\label{sec:RadioL}

We assume that the initial electron energy distribution when injected into the lobe is a power law in energy given by:
\begin{equation}
n(\gamma_{\rm i},t_{\rm i})d\gamma_{\rm i}=n_0\gamma_{\rm i}^{-p}d\gamma_{\rm i}
\label{eq:ni}
\end{equation}
with, for case [A], $p=2.14$ as in \cite{1997MNRAS.292..723K} and $\gamma_{\rm i}$ ranging between $\gamma_{\rm min}=1$ and $\gamma_{\rm max}=10^6$ (Lorentz factors of $\gamma\sim10^3$ are required to produce upscattering of the CMB in the X-ray and Lorentz factors of $\gamma\geq 10^4$ are needed for GHz synchrotron radiation in the radio for typical magnetic field strengths). The constant $n_0$ is found by integration:
\begin{equation}
n_0=\frac{u_{\rm e}(t_{\rm i})}{m_{\rm e}c^2}\left(\displaystyle\int_{\gamma_{\rm min}}^{\gamma_{\rm max}}(\gamma_{\rm i}-1)\gamma_{\rm i}^{-p}\,d\gamma_{\rm i}\right)^{-1}.
\label{eq:n0}
\end{equation}
Increasing $\gamma_{\rm max}$ has little effect on the luminosity. However, $p$ can range from $2$ to $3$ \citep{1987MNRAS.225....1A} and the $P$--$\alpha$ correlation  of \cite{1999AJ....117..677B} suggests sources with higher jet power have higher $p$. Also, $\gamma_{\rm min}$, often assumed to be $1$ in previous models, may in fact be higher and influence the luminosities, an issue we explore in \S~\ref{sec:params}. The minimum injected Lorentz factor $\gamma_{\rm min}$ has taken various values from $1$ to $10^4$ for describing individual sources \citep{2006ApJ...644L..13B}.

We also assume
\begin{equation}
u_{\rm B}(t)=\frac{rp_{\rm l}(t)}{(\Gamma_{\rm l}-1)(r+1)}
\label{eq:uB}
\end{equation}
and
\begin{equation}
u_{\rm e}(t)=u_{\rm B}(t)/r,
\label{eq:ue}
\end{equation}
that is, the ratio of the energy density in the particles to that in the magnetic field is a constant $r=(1 + p)/4$, based on minimum energy arguments, adopted from \cite{1997MNRAS.292..723K}.

We now describe the determination of $\gamma_{\rm i}$ and $d\gamma_{\rm i} / d\gamma$. First
\begin{equation}
\label{eq:An0}
\frac{1}{V_{\rm l}}\frac{dV_{\rm l}}{dt}=
\begin{cases}
\frac{a_1}{t} & t<t_{\rm j} \\
\frac{3a_2}{t-t_{\rm j}+C/\sqrt{k}} & t>t_{\rm j} \\ 
\end{cases}
\end{equation}
where $a_1=1-(4+\beta)/[\Gamma_{\rm_c}(5-\beta)]$ and $a_2=2/(3\Gamma_{\rm l}+2-\beta)$.
Equation~(\ref{eq:loss}) may be integrated to yield:
\begin{equation}
\label{eq:An1}
\frac{F(t)}{\gamma}-\frac{F(t_{\rm i})}{\gamma_{\rm i}}=A(t,t_{\rm i})
\end{equation}
where
\begin{equation}
\label{eq:An2}
A(t,t_{\rm i})=\frac{4}{3}\frac{\sigma_{\rm T}}{m_{\rm e} c}\displaystyle\int_{t_{\rm i}}^{t} (u_{\rm B}(t^\prime)+u_{\rm C})F(t^\prime)\,dt^\prime
\end{equation}
and
\begin{equation}
\label{eq:An3}
F(t)=\begin{cases}
F_1(t)\equiv t^{-a_1/3} & t<t_{\rm j} \\
F_2(t)\equiv (t-t_{\rm j}+C/\sqrt{k})^{-a_2} & t>t_{\rm j} \\ 
\end{cases}.
\end{equation}
The integral in $A(t,t_{\rm i})$ (equation~(\ref{eq:An2})) has an analytic form as the integrand is proportional to a power of $t$ for $t<t_{\rm j}$ and a power of $(t-t_{\rm j}+C/\sqrt{k})$ for $t>t_{\rm j}$.
Thus for $t_{\rm i}\leq t \leq t_{\rm j}$
\begin{equation}
\label{eq:An4}
\gamma_{\rm i}=\frac{\gamma F_1(t_{\rm i})}{F_1(t)-\gamma A(t,t_{\rm i})}
\end{equation}
and for $t_{\rm i}\leq t_{\rm j}\leq t$,
\begin{equation}
\label{eq:An5}
\gamma_{\rm i}=\frac{\gamma_{t_{\rm j}} F_1(t_{\rm i})}{F_1(t_{\rm j})-\gamma_{t_{\rm j}}  A(t_{\rm j},t_{\rm i})}
\end{equation}
with
\begin{equation}
\label{eq:An6}
\gamma_{t_{\rm j}}=\frac{\gamma F_2(t_{\rm j})}{F_2(t)-\gamma A(t,t_{\rm j})}.
\end{equation}
Then $d\gamma_{\rm i} / d\gamma$ can be found by differentiating equations~(\ref{eq:An4}) and (\ref{eq:An5}).

The total radio power at a frequency $\nu$ can be then found by an integration over time:
\begin{equation}
P_{\nu}(t)=\displaystyle\int_0^{\min[t,t_{\rm j}]}\,dP_\nu
\label{eq:Pv}
\end{equation}
where $d P_\nu$ is given by equation~(\ref{eq:synchrotron}).

\subsection{Inverse-Compton X-ray luminosity from the lobe}\label{sec:IC}

We calculate the IC emission in the X-ray as a function of time as in \cite{2010MNRAS.tmp.1004N}. An electron with a Lorentz factor of $\gamma$ boosts a CMB photon with frequency $\nu_{\rm CMB}$ to frequency $\gamma^2\nu_{\rm CMB}$. We can make a simplification that all CMB photons are at the peak CMB frequency $5.879\times 10^{10}~{\rm Hz}~{\rm K}^{-1}\times 2.7~{\rm K}~(1+z)$.
Then, the IC power at a frequency $\nu$ is estimated as
\begin{equation}
P_{\nu}(t)=\displaystyle\int_0^{\min[t,t_{\rm j}]}\frac{1}{6\pi}\sigma_{\rm T} c u_{\rm c} \frac{\gamma^3}{\nu}n(\gamma)\,dV.
\label{eq:Pic}
\end{equation}
This is a simplification and we are not considering the full spectrum of CMB photons but \cite{2010MNRAS.tmp.1004N} shows that it is a reasonable measure of the X-ray luminosity, correct to within an accuracy of $60$~per~cent.

\subsection{Evolutionary tracks}\label{sec:evo}

\begin{figure}
\centering
\includegraphics[width=0.47\textwidth]{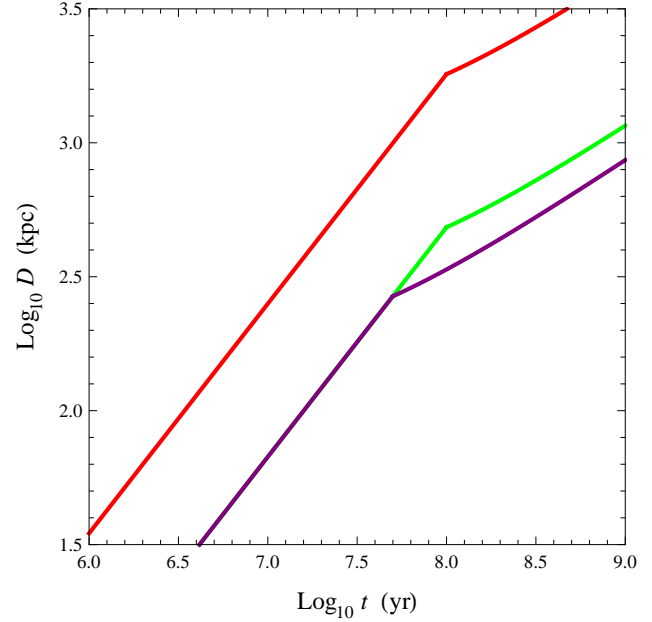}
\caption{Growth in the length of a lobe as a function of time.
$Q_{\rm j}=1.5\times 10^{40}~{\rm W}$, $t_{\rm j}=10^8~{\rm yr}$ (top/red), 
$Q_{\rm j}=1.5\times 10^{38}~{\rm W}$, $t_{\rm j}=10^8~{\rm yr}$ (middle/green), 
$Q_{\rm j}=1.5\times 10^{38}~{\rm W}$, $t_{\rm j}=5\times10^7~{\rm yr}$ (bottom/purple). The growth rate of the lobe length changes once the jets turn off.}
\label{fig:l}
\end{figure}

\begin{figure}
\centering
\includegraphics[width=0.47\textwidth]{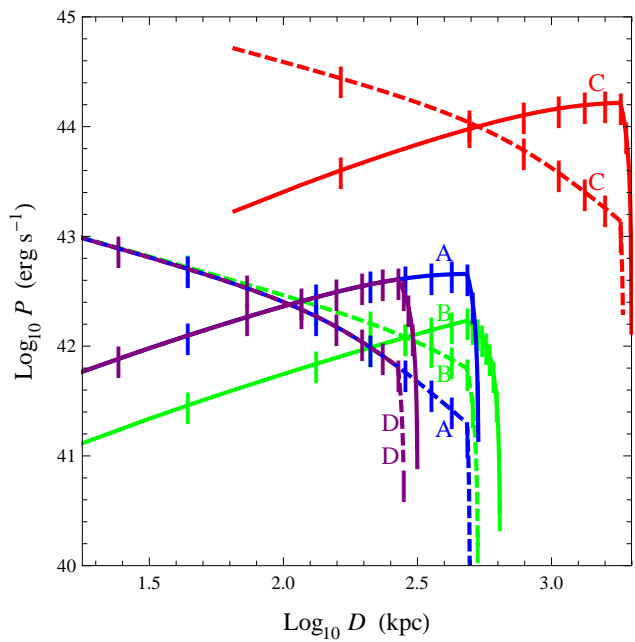}
\caption{Power-linear size ($P$-$D$) diagram for the evolution of the source in the X-ray ($1~{\rm keV}$) and the radio ($151~{\rm MHz}$). Radio luminosities are dashed, X-ray luminosities are solid. Tick marks represent time intervals of $0.2t_{\rm j}$ increments, starting from $0.2t_{\rm j}$.
[A/blue] $Q_{\rm j}=1.5\times 10^{38}~{\rm W}$, $t_{\rm j}=10^8~{\rm yr}$, $z=2$,
[B/green] $Q_{\rm j}=1.5\times 10^{38}~{\rm W}$, $t_{\rm j}=10^8~{\rm yr}$, $z=1$,
[C/red] $Q_{\rm j}=1.5\times 10^{40}~{\rm W}$, $t_{\rm j}=10^8~{\rm yr}$, $z=2$,
[D/purple] $Q_{\rm j}=1.5\times 10^{38}~{\rm W}$, $t_{\rm j}=5\times10^7~{\rm yr}$, $z=2$.}
\label{fig:lr}
\end{figure}

\begin{figure}
\centering
\includegraphics[width=0.47\textwidth]{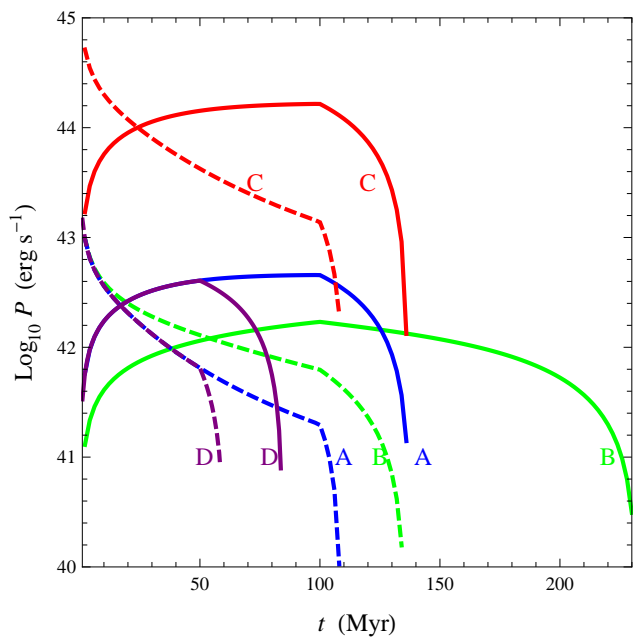}
\caption{Power-time ($P$-$t$) diagram for the evolution of the source in the X-ray ($1~{\rm keV}$) and the radio ($151~{\rm MHz}$). Radio luminosities are dashed, X-ray luminosities are solid. 
[A/blue] $Q_{\rm j}=1.5\times 10^{38}~{\rm W}$, $t_{\rm j}=10^8~{\rm yr}$, $z=2$,
[B/green] $Q_{\rm j}=1.5\times 10^{38}~{\rm W}$, $t_{\rm j}=10^8~{\rm yr}$, $z=1$,
[C/red] $Q_{\rm j}=1.5\times 10^{40}~{\rm W}$, $t_{\rm j}=10^8~{\rm yr}$, $z=2$,
[D/purple] $Q_{\rm j}=1.5\times 10^{38}~{\rm W}$, $t_{\rm j}=5\times10^7~{\rm yr}$, $z=2$.}
\label{fig:lt}
\end{figure}

We plot the time evolution of the length of the lobe in Figure~\ref{fig:l}. After time $t_{\rm j}$ the length expands slower than when the jet was on. Power-linear size ($P$-$D$) diagrams for various source parameters are presented in Figure~\ref{fig:lr}. The figure is useful for comparing our model to the models of
\cite{1997MNRAS.292..723K}, \cite{1999AJ....117..677B} and \cite{2010MNRAS.tmp.1004N} which all present $P$-$D$ diagrams.
For the goals of this paper we simply need, however, the time evolution of the X-ray and radio luminosities. We will consider the X-ray luminosity at $1~{\rm keV}$ and the radio luminosity at $151~{\rm MHz}$. The time evolution at these two frequencies is presented in Figure~\ref{fig:lt}.

\begin{figure}
\centering
\includegraphics[width=0.47\textwidth]{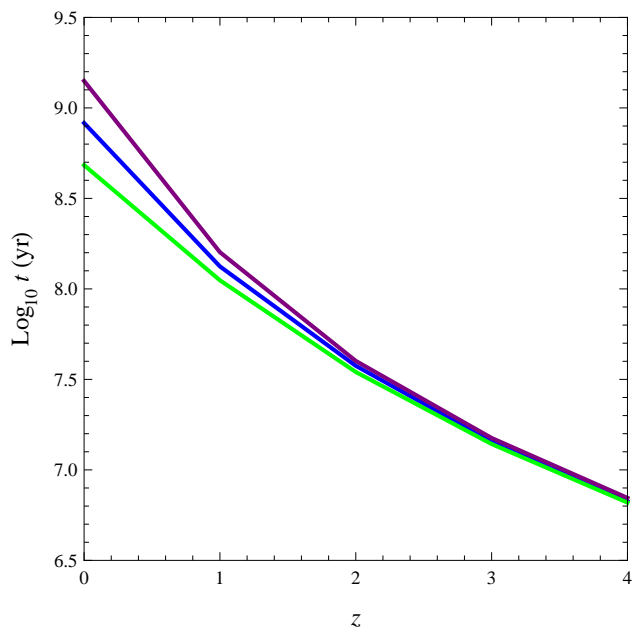}
\caption{The length of time a source continues to radiate in the X-ray ($1~{\rm keV}$) after the jet has turned off as a function of redshift (similar for all jet energies). Jet lifetimes of $t_{\rm j}=5\times10^7~{\rm yr}$, $10^8~{\rm yr}$ and $5\times10^8~{\rm yr}$ (bottom, middle, top / green, blue, purple) are used for the figure.}
\label{fig:go}
\end{figure}

In Figure~\ref{fig:lt} we see that while the jet is switched on, the radio luminosity declines and the X-ray luminosity rises. After the time $t_{\rm j}$ when the jet turns off the radio luminosity precipitates very quickly and the X-ray luminosity falls but at a less rapid rate than this. At higher redshifts, the X-ray luminosity is stronger due to the $(1+z)^4$
rise in the energy density of the CMB. The luminosities also fall more rapidly at higher redshifts due to increased IC losses from the CMB. 
In Figure~\ref{fig:go} we present the length of time a radio source continues to radiate in the X-ray after jet activity has ceased. At $z=2$, a source will radiate in the X-ray for about $30~{\rm Myr}$ after the jet activity stops at $t_{\rm j}=10^8~{\rm yr}$, since there is no new injection of particles and existing particles had their Lorentz factors reduced to less than $\gamma\sim10^3$. 

\subsection{Model parameters}\label{sec:modpar}

Model parameter case [A] has $t_{\rm j}=10^8~{\rm yr}$, $p=2.14$, $\gamma_{\rm min}=1$, $\rho_0= 1.67\times 10^{-23}~{\rm kg}~{\rm m}^{-3}$ and $\beta=1.5$.
These parameters are inferred from observations and similar to parameters used in previous models for radio sources. 
Parameters such as the minimum injected Lorentz factor, $\gamma_{\rm min}=1$, however, are not well constrained by observations and may range from $1$ to $10^4$ \citep{2006ApJ...644L..13B}.
Other sets of model parameters we investigate (\S~\ref{sec:params}) are:
[B] $t_{\rm j}\to 5\times10^7~{\rm yr}$,
[C] $t_{\rm j}\to 5\times10^8~{\rm yr}$,
[D] $p\to 3$,
[E] $\gamma_{\rm min}\to 2000$,
[F] $p\to 3$ and $\gamma_{\rm min}\to 2000$,
[G] $p$ correlated with $Q_{\rm j}$ (see \S~\ref{sec:params}),
[H] $\rho_0\to 1.67\times 10^{-24}~{\rm kg}~{\rm m}^{-3}$,
[I] $\beta\to 2$
[J] birth function (see \S~\ref{sec:BF}) $\Delta z\to 1$,
[K] jet-power power-law $-2.6\to -3.0$, and
[L] jet-power power-law $-2.6\to -2.0$.

\subsection{Comparison with other models in literature}\label{sec:lit}

Relevant earlier work in the subject goes back to 
\cite{1989MNRAS.239..173G},  which produced the first
quantitative study of the evolution of old radio lobes and their fading via
IC processes. A number of models since have been developed for the evolution of the size and luminosity of radio lobes.

The model presented in \cite{1997MNRAS.286..215K} and \cite{1997MNRAS.292..723K} provides an analytical description of the evolution of FR II radio sources that includes adiabatic, synchrotron, and IC losses on the source's luminosity. The model assumes a self-similar expansion of the jet and lobes, with a constant axial ratio throughout evolution. The evolution does not go beyond the lifetime of the jet. 

To investigate spectral dependencies \cite{1999AJ....117..677B} created a model based on that of \cite{1997MNRAS.286..215K} and \cite{1997MNRAS.292..723K}, however, their model considers a more complex injection into the lobe governed by breaks in the energy distribution determined by the dwell time and magnetic field strengths in the hotspot, rather than assuming a constant injection index. The model also considers the hotspot pressure rather than the lobe pressure to govern adiabatic losses out of the hotspot in order to find the lobe luminosity. Axial ratios in this model increase with time (which correctly accounts for the observed larger increase for more powerful jets), as the hotspot pressure does not scale in the same way as the lobe pressure. 
This model better reproduces observations of jets, such as in \cite{1989MNRAS.239..401L} and  \cite{1984MNRAS.210..929L}, which found that axial ratios are larger in sources with higher jet powers.

\cite{2002A&A...391..127M} extend the model of  \cite{1999AJ....117..677B}.
Instead of specifying the electron distribution that
enters the lobe, the model accounts for electrons accelerating by the first-order Fermi process at the termination shock and
then propagating through the hotspot into the lobe. To match $P$-$D$ observations, a reacceleration
process during propagation through the head is included which compensates for the adiabatic losses between the termination shock and the lobe.

\cite{2006MNRAS.372..381B} and 
\cite{2008MNRAS.388..677W} compare the three models of \cite{1997MNRAS.286..215K},  \cite{1999AJ....117..677B}, and \cite{1999AJ....117..677B}.
Monte Carlo simulations for each model are used to predict 
radio powers, sizes, redshifts and spectral
indices of an artificial sample, which are compared to the data of the low-frequency radio survey 
3CRR, 6CE and 7CRS. \cite{2006MNRAS.372..381B} find that no existing
model can give acceptable fits to all the properties of the surveys considered, and the simplest \cite{1997MNRAS.286..215K} model is somewhat better at fitting the data.
\cite{2008MNRAS.388..677W} evolve some of the properties of FR II sources with redshift to have the artificial samples to fit the observations, and also finds \cite{1997MNRAS.286..215K} best matches observations.

The model of \cite{2010MNRAS.tmp.1004N} is the first to go beyond the stopping of the jet after a time $t_{\rm j}$ and consider the X-ray emission of the lobes from IC scattering of the CMB photons. The model is a variant of \cite{1997MNRAS.286..215K} and \cite{1997MNRAS.292..723K}, also assuming a constant axial ratio. The model does not evolve simply according to the self-similar evolution of lobe length determined by the characteristic length scale.

The model in the present work also goes beyond the cessation of the jet in the formalism of \cite{2010MNRAS.tmp.1004N} and considers the growth of the lobe as described by \cite{1999AJ....117..677B} where a compact hotspot with a pressure distinct from the lobe pressure that determines lobe length growth feeds the lobes, and in which axial ratios grow over time during jet activity.  The model is also analytic rather than numeric.
We are able to go beyond the self-similarity expansion after the jet stops (i.e. no longer rely on a characteristic length scale).
We are not required to fix a constant axial ratio during jet activity and solve the pairs of differential equations as in equation~(6) of \cite{2010MNRAS.tmp.1004N} to solve for the evolution of length and pressure, which results in the volume evaluated as $V_{\rm l}(t)=\pi L_{\rm j}(t)^3/(4R^2)$ not precisely agreeing with the volume of the lobe determined by integration of volume elements (equation~(8) of \cite{2010MNRAS.tmp.1004N}).

\section{Observationally inferred distribution parameters}\label{sec:OIDP}

In this section we discuss the radio luminosity function (RLF) of Fanaroff and Riley Class II radio sources, the empirically-inferred birth function of such sources and the empirically-inferred distribution of jet energies which are necessary to find the distribution of the IC ghosts and the completely dark, i.e. non-observable, radio and X-ray lobes.
An IC ghost refers to a galaxy that has its jets turned off but still emits X-ray radiation due to the upscattering of CMB photons. 
An {\it observable} IC ghost is the term we use to call an IC ghost that radiates above a given X-ray flux limit (but lacks detectable synchrotron radiation).
In \S~\ref{sec:OIDP}-\ref{sec:XLF} we use the normative model parameters of case~[A]. In \S~\ref{sec:params} we explore model parameter cases [B]-[L].

\subsection{RLF of FR II sources}\label{sec:RLF}

We will use the RLF of high radio power FR II sources determined by \cite{2001MNRAS.322..536W} from the 7CRS, 6CE and 3CRR samples, to aid in predicting the density of radio lobes including IC ghosts and completely non-observable ones. The radio luminosity function describes the space density per unit comoving volume of sources as a function of luminosity as derived from the surveys.
The luminosity function was converted from the cosmology $H_0=50~{\rm km}~{\rm s}^{-1}~{\rm Mpc}^{-1}$, $\Omega_{\rm M}=0$, $\Omega_\Lambda=0$, $\Omega_{\rm k}=1$ to the cosmology we use in this paper with the relation \citep{1985MNRAS.217..601P}
\begin{equation}
\rho_1(P_1,z)\frac{dV_1}{dz}=\rho_2(P_2,z)\frac{dV_2}{dz}
\end{equation}
where the indices refer to a cosmological model, $P_{\rm i}$ is the luminosity derived from the flux density and redshift in model $i$ and $V_{\rm i}$ is comoving volume.

The RLF of \cite{2001MNRAS.322..536W} will reflect only those radio galaxies that are above the flux limit (at low radio frequencies $\sim151~{\rm MHz}$) of $0.5~{\rm Jy}$ \citep{2001ApJ...560L.115G}, which is the lowest flux limit of the surveys they used to determine the RLF.

\subsection{Birth function of radio galaxies}\label{sec:BF}

We assume the empirically-inferred birth function of radio sources given by \cite{1999AJ....117..677B}:
\begin{equation}
p(z)dz\propto e^{-\frac{1}{2}\left(\frac{z-z_1}{\Delta z}\right)^2}dz
\end{equation}
with $z_1=2.2$ and $\Delta z=0.6$.
Most sources are born during the quasar era, $z=1.5$--$3$. By a ``birth'' we mean jet activity initiates. The birth function was determined by \cite{1999AJ....117..677B} such that simulations of sources best match 3C and 7C data.

\begin{figure}
\centering
\includegraphics[width=0.47\textwidth]{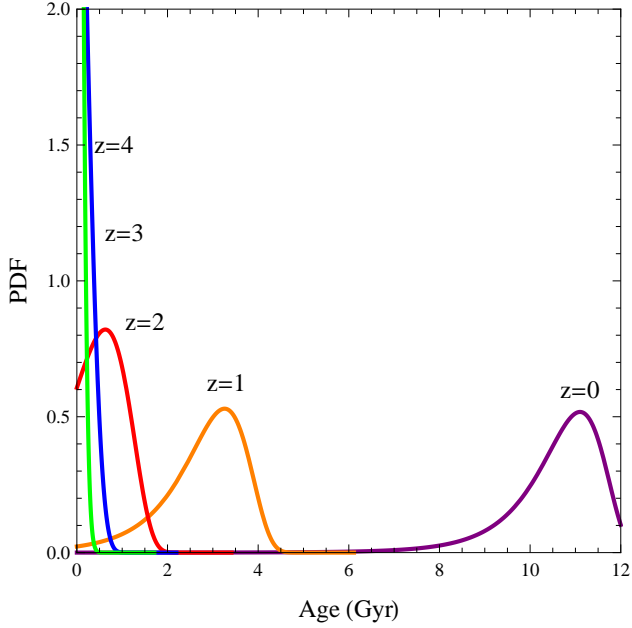}
\caption{Probability density function (PDF) for the age of radio sources at various redshifts. $z=0,1,2,3,4$ are shown. The PDF of ages is useful for determining the density of IC ghosts as well as completely dark X-ray and radio lobes.}
\label{fig:pAge}
\end{figure}

It is useful to convert the birth function to the probability density function (PDF) for the age of radio sources as a function of redshift. This is accomplished through the equation
\begin{equation}
p_{\rm A}(t,z) dt = \frac{p(z_{\rm t})}{\displaystyle\int_z^\infty p(z^\prime) \,dz^\prime} dz
\end{equation}
where $p_{\rm A}(t,z)$ gives the probability density that a source is age $t$ at redshift $z$
 and $z_{\rm t}$ is the redshift of a source at its birth for it to appear age $t$ when observed at a redshift of $z$.
The PDF for the age of radio sources at $z=0,1,2,3,4$ is presented in Figure~\ref{fig:pAge}.

\subsection{Distribution of jet powers}\label{sec:JP}

We use the  empirically-inferred probability density function of jet energies by \cite{1999AJ....117..677B}. The distribution is given by
\begin{equation}
p(Q)dQ\propto Q^{-2.6} 
\end{equation}
with $Q_{\rm min}=5\times 10^{37}~{\rm W}\leq Q\leq 5\times 10^{42}~{\rm W}=Q_{\rm max}$. The lower-energy jets are most abundant according to this power law. At high redshifts we may be only seeing the brighter sources (those that are younger and/or have higher jet power) above the radio flux limit, meaning that a large number of radio sources may go undetected.
We will also consider modifying the power-law index from $-2.6$ to $-2$ and $-3$.
\cite{2008MNRAS.388..677W} find that a power-law index of $-2$ better matches observed $P-D-z$ and $\alpha$ distributions.
\cite{2006MNRAS.372..381B} on the other hand find that a steeper power-law index of $-3$ is preferable.

\section{Observable distribution of inverse-Compton ghosts and active sources}\label{sec:CG}

\begin{figure}
\centering
\includegraphics[width=0.47\textwidth]{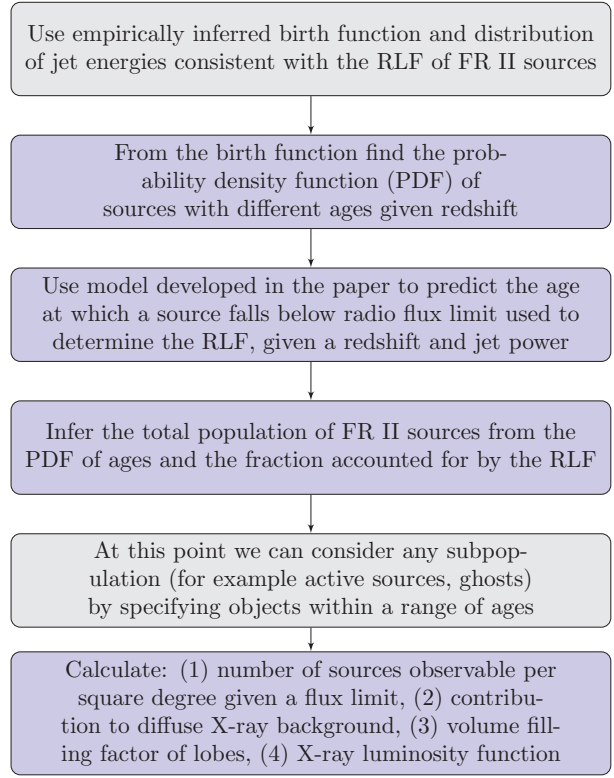}
\caption{A flowchart summarizing the major steps in the analysis of \S~\ref{sec:OIDP} and \ref{sec:CG}.}
\label{fig:flowchart}
\end{figure}

We are interested in the distribution of all radio galaxies that have ongoing jet activity, the IC ghosts and galaxies with dead radio lobes that no longer exhibit detectable radiation.
Dead lobes refer to lobes that are completely X-ray and radio dark. Although undetectable, the fossil radio lobes are still an important component of the intergalactic medium: \cite{2001A&A...366...26E} demonstrated that radio plasma in the lobes of galaxies, after jet activity has ceased, can be revived even up to $2$~Gyr later by compression in a shock wave produced by large-scale structure formation.
The major steps of the analysis are summarized in a flowchart  in  Figure~\ref{fig:flowchart}.

The ratio of IC ghosts to radio sources with jets on at a given redshift $z$ is:
\begin{equation}
\frac{{\rm IC\,ghosts}}{{\rm sources\,with\,jets\,on}}=\frac{\displaystyle\int_{t_{\rm j}}^{t_{\rm max}}p_{\rm A}(t,z) \,dt}{\displaystyle\int^{t_{\rm j}}_0 p_{\rm A}(t,z) \,dt}
\label{eq:ratioGF}
\end{equation}
where $t_{\rm j}$ is the jet lifetime and $t_{\rm max}$ is the time at which the X-ray luminosity goes to $0$, or, if we care about the observable distribution of the ghosts, then the time at which the X-ray luminosity drops below some flux limit. We may alter the integration limits in the numerator accordingly to derive the ratio of sources with dead lobes to sources with jet activity (specifically, integrate from $t_{\rm max}$ to $\infty$).

Thus, if we have an accurate idea of the total density of radio sources with current jet activity at each redshift, then we can estimate the density of IC ghosts and dead lobes as well using a ratio as in equation~(\ref{eq:ratioGF}), with appropriate integration limits.

We can use the RLF of FR II objects described in \S~\ref{sec:RLF} to find the total density of radio sources with jets on, $\rho_{\rm sjo}(z)$. Recall that the RLF will only account for those radio galaxies that are above the flux limit of the radio survey.
In other words, because the radio luminosity of a source decreases with time in our model, the RLF only accounts for galaxies that are younger than some time $t^{\prime}(Q_{\rm j},z)\leq t_{\rm j}$ at each redshift in the evolution of the radio source with jet power $Q_{\rm j}$ and redshift $z$. Then, to find the total density of radio galaxies with jet activity at a given redshift $z$, the ratio
\begin{equation}
\frac{{\rm sources\,with\,jets\,on}}{{\rm FR\,II\,from\,RLF}}=\frac{\displaystyle\int_{0}^{t_{\rm j}}p_{\rm A}(t) \,dt}{\displaystyle\int_{Q_{j,{\rm min}}}^{Q_{j,{\rm max}}}p(Q)\displaystyle\int^{t^{\prime}(Q,z)}_0 p_{\rm A}(t) \,dt\,dQ}
\label{eq:ratioFR II}
\end{equation}
is used.
Here we are averaging over the empirically-inferred jet powers. The density of FR II objects accounted for by the RLF is found by a simple integration of the RLF over luminosity. Multiplying this density by the ratio in equation~(\ref{eq:ratioFR II}) gives $\rho_{\rm sjo}(z)$ that was sought. Then, at any redshift, the density of {\it all}
objects active or dead, $\rho_{\rm tot}(z)$, is:
\begin{equation}
\rho_{\rm tot}(z)=\frac{\rho_{\rm sjo}(z)}{\displaystyle\int_0^{t_{\rm j}} p_{\rm A}(t,z) \,dt}.
\label{eq:ratiotot}
\end{equation}

An X-ray limiting flux of 
$3.0\times10^{-19}~{\rm J}~{\rm m}^{-2}~{\rm s}^{-1}$ will be considered, similar to that of {\it Chandra} X-ray Deep Field North (CDFN) survey.
Also, a radio limiting flux density of $7.55\times10^{-19}~{\rm J}~{\rm m}^{-2}~{\rm s}^{-1}$ will be considered, corresponding to the flux limit of the 7CRS survey which had the lowest flux limit of the surveys used in the determination of the RLF of FR II sources.

We are interested in the number of X-ray ghosts above the X-ray flux limit per square degree.
This is obtained by the equation:
\begin{equation}
N=\left(4\pi\left(\frac{180}{\pi}\right)^2\right)^{-1}\displaystyle\int \rho_{{\rm ghosts } > {\rm x.f.l.}}\,dV_{\rm c} 
\label{eq:number density}
\end{equation}
where $\rho_{{\rm ghosts } > {\rm x.f.l.}}$ is the comoving number density of the IC ghosts above the X-ray flux limit obtained from multiplying $\rho_{\rm sjo}$ and the ratio in equation~(\ref{eq:ratioGF}).
\begin{equation}
dV_{\rm c}=4\pi \frac{c}{H_0}\frac{D_{\rm L}(z)^2}{(1+z)^2 E(z)}dz
\label{eq:comoving}
\end{equation}
 is the comoving volume of a shell at $z$ of thickness $dz$, where $D_{\rm L}$ is luminosity distance and 
\begin{equation} 
E(z)=\left(\Omega_{\rm M}(1 + z)^3 + \Omega_\Lambda+\Omega_{\rm k}(1+z)^2\right)^{1/2}.
\label{eq:E}
\end{equation} 
(see \cite{2001ApL&C..40..161C}).
We integrate from $z=0$ to $z=4$ to find the number density. 

We predict $\sim29$ powerful double-lobed galaxies per square degree visible in the X-ray above the X-ray flux limit (of the CDFN survey), of which $\sim4$ (approximately $13$~per~cent) are IC ghosts.
Our prediction is $17$~per~cent of the number density of extended X-ray sources estimated by the CDFN survey: \cite{2002AJ....123.1163B} find a surface density of $167^{+97}_{-67}$ ($1\,\sigma$) extended X-ray sources (including X-ray clusters) at a limiting soft-band flux of $3\times10^{-16}~{\rm erg}~{\rm cm}^{-2}~{\rm s}^{-1}=3.0\times10^{-19}~{\rm J}~{\rm m}^{-2}~{\rm s}^{-1}$ inferred from the $\sim1~{\rm Ms}$ CDFN observation.  

\begin{figure}
\centering
\includegraphics[width=0.47\textwidth]{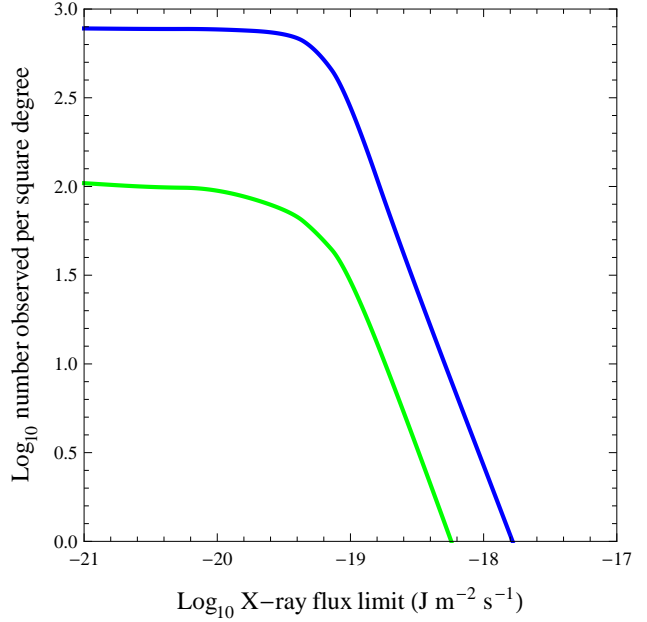}
\caption{The number per square degree of all radio lobes visible in the X-ray (top/blue) and the number of observable IC ghosts (bottom/green) as a function of X-ray flux limit.}
\label{fig:no}
\end{figure}

We also calculate the expected number density of sources for an X-ray flux limit of $2\times10^{-15}~{\rm erg}~{\rm cm}^{-2}~{\rm s}^{-1}=2\times10^{-18}~{\rm J}~{\rm m}^{-2}~{\rm s}^{-1}$, which was the depth of the Subaru-XMM Deep Field North observation. \cite{2010MNRAS.403.2063F} identify $57$ X-ray cluster candidates and $6$ sources possibly due to X-ray emission of radio lobes in the Subaru-XMM Deep Field which surveyed a $1.3$ square degree region (about $5$ sources per square degree). Our model predicts $1$ X-ray source per square degree above the flux limit. The predicted number of observable IC ghosts and total radio lobes per square degree in the X-ray as a function of the X-ray flux limit is presented in Figure~\ref{fig:no}.

\begin{figure}
\centering
\includegraphics[width=0.47\textwidth]{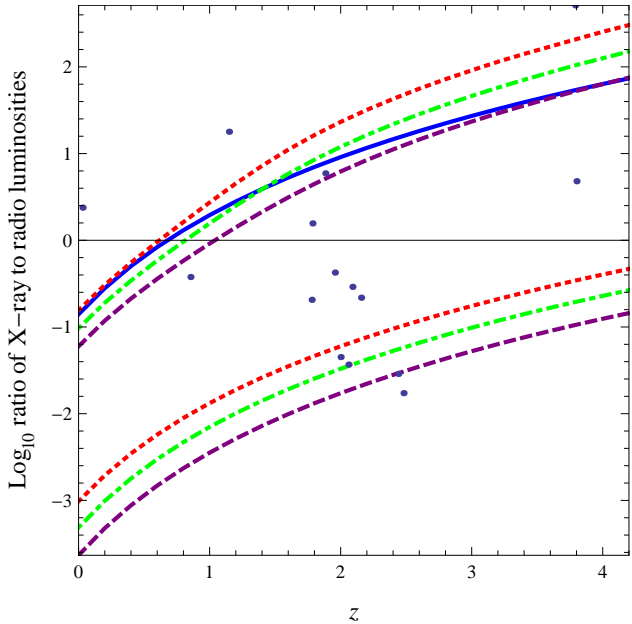}
\caption{X-ray ($1~{\rm keV}$) to radio ($151~{\rm MHz}$) luminosity ratios as a function of redshift. A jet lifetime of $t_{\rm j}=10^8~{\rm yr}$ is used for the figure. The ratios in our model at $t=0.05t_{\rm j}$ (bottom set of lines) and $t=t_{\rm j}$ (top set) are shown for 
[1] $Q_{\rm j}=1.5\times 10^{38}~{\rm W}$ (dotted/red),
[2] $Q_{\rm j}=1.5\times 10^{40}~{\rm W}$ (dot-dashed/green),
[3] $Q_{\rm j}=1.5\times 10^{42}~{\rm W}$ (dashed/purple). Once the jet turns off, radio luminosity drops rapidly and the ratio tends to infinity.
The time-averaged ratio ($0.14(1+z)^{3.8}$) over the duration in which the sources were both X-ray and radio bright in the rest frame calculated in Nath (2010) is also shown (solid/blue).
The blue points represent actual sources (see \S~\ref{sec:CG}).
}
\label{fig:LxLr}
\end{figure}

Also of interest are the luminosity ratios of the X-ray ($1~{\rm keV}$) to radio ($151~{\rm MHz}$) predicted by our model. \cite{2004MNRAS.353..523C} use this ratio to convert the RLF of FR II objects to an X-ray luminosity function (the ratio is taken to be a conservative value of $1$). \cite{2010MNRAS.tmp.1004N} develops a model for the evolution of the X-ray and radio luminosities of double-lobed radio galaxies and
takes the time-averaged ratio over the duration in which the sources were both X-ray and radio bright (in the rest frame) to also obtain an XLF for extended X-ray sources.
We plot the ratios at $t=0.05t_{\rm j}$ and $t=t_{\rm j}$ of the jet as a function of redshift in Figure~\ref{fig:LxLr} for various jet energies. For comparison, the time-averaged ratio $0.14(1+z)^{3.8}$ of \cite{2010MNRAS.tmp.1004N} is included as well,
which \cite{2010MNRAS.tmp.1004N} obtains 
by averaging the ratio of luminosities in X-ray and radio
frequencies of that paper's model over a time until the X-ray or radio emission
drops rapidly.
We also plot ratios obtained from real observations from 
4C\,60.07 ($z=3.79$) \citep{2009ApJ...702L.114S},
PKS\,1138-262 ($z=2.156$) \citep{2002ApJ...567..781C},
PKS\,0156-252 ($z=2.09$),
PKS\,0406-244 ($z=2.44$),
PKS\,2036-254 ($z=2.00$),
PKS\,2048-272 ($z=2.06$) \citep{2005A&A...433...87O},
3C\,432 ($z=1.785$),
3C\,294 ($z=1.779$),
3C\,191 ($z=1.956$) \citep{2006MNRAS.371...29E},
6C\,0905+39 ($z=1.883$) \citep{2008MNRAS.386.1774E},
MRC\,2216-206 ($z=1.148$),
MRC\,0947-249 ($z=0.854$) \citep{2010MNRAS.401.1500L},
4C\,23.56 ($z=2.48$) \citep{2007MNRAS.376..151J},
4C\,41.17 ($z=3.8$) \citep{2003ApJ...596..105S} and
3C\,98 ($z=0.0306$) \citep{2005ApJ...632..781I}. The extended X-ray lobe luminosities ($1~{\rm keV}$ rest frame) are obtained from the mentioned papers (recalculated to our cosmology) and the radio luminosities are estimated from NASA/IPAC Extragalactic Database (NED) photometric data. The size of the sample is $15$, but we can already see large scatter in the ratios. Our model is also able to predict a wide range of ratios. Some sources however appear to be above the ranges predicted by our model. We are only showing the range of ratios while the jet is on. Once the jet turns off the radio luminosity plummets very rapidly and the X-ray to radio ratio tends towards infinity. It may be possible that some of the sources with high ratios have their jets turned off recently. Or, it may be possible that the sources with high X-ray to radio luminosity ratios have jet lifetimes greater than $10^8~{\rm yr}$ or that environment or injection spectrum parameters of these sources deviate significantly from those of case [A].
We can obtain higher ratios if we increase $p$ or lower the surrounding density. Also, if $\beta=2$, the value used in \cite{2010MNRAS.tmp.1004N}, the ratios are found to not depend on jet power.
We are likely underestimating the X-ray to radio luminosity ratios of the $15$ sources because the NED radio luminosity includes the lobes, hotspot and nucleus rather than just the lobes.

\subsection{Diffuse X-ray background contribution}\label{sec:DXR}

The diffuse X-ray background contribution from accumulated Compton lobes and ghosts is also calculated.
It is important to check that this flux does not exceed the observed CXB.
The diffuse X-ray flux (in ${\rm J}~{\rm m}^{-2}~{\rm s}^{-1}~{\rm deg}^{-2}$) can be found by an integration across comoving volume shells and hence $z$. 
\begin{equation}
F=\left(4\pi\left(\frac{180}{\pi}\right)^2\right)^{-1}\displaystyle\int f(t,z)
\rho_{\rm tot}(z)
\,dV_{\rm c} 
\end{equation}
with
\begin{equation}
f(t,z)\equiv \displaystyle\int_{t_{\rm j}}^{t_{\rm max}} \frac{2 L_{\rm x}(t)}{4 \pi D_{\rm L}^2(z)} \times p_{\rm A}(t,z) \,dt,
\end{equation}
for example, gives the X-ray flux of the IC ghosts. 
The luminosity will depend on jet power $Q_{\rm j}$, so we bin jet power, calculate $F$ at each and take a weighted average according to the empirically-inferred distribution of jet powers. We calculate luminosities at $1$~keV observed energies.

From studies of the CXB by \cite{2007ApJ...661L.117H} and \cite{2006ApJ...645...95H}, the flux density of the CXB (after all resolved X-ray point and extended sources have been removed) at $\sim1~{\rm keV}$ is approximately $4\times10^{-12}~{\rm erg}~{\rm cm}^{-2}~{\rm s}^{-1}~{\rm deg}^{-2} =4\times10^{-15}~{\rm J}~{\rm m}^{-2}~{\rm s}^{-1}~{\rm deg}^{-2}$.
The X-ray flux from {\it all} the sources predicted by our model is $9.7\times 10^{-17}~{\rm J}~{\rm m}^{-2}~{\rm s}^{-1}~{\rm deg}^{-2}$, about $2.5$~per~cent of the observed but unresolved CXB. Of course, many of the objects in the calculation of X-ray flux are above the flux limit and should be excluded. The flux from all extended sources {\it below} the X-ray flux limit is $7.3\times 10^{-17}~{\rm J}~{\rm m}^{-2}~{\rm s}^{-1}~{\rm deg}^{-2}$.
Compton ghosts contribute about a tenth of the total X-ray flux: $1.1\times 10^{-17}~{\rm J}~{\rm m}^{-2}~{\rm s}^{-1}~{\rm deg}^{-2}$, of which $3.9\times 10^{-18}~{\rm J}~{\rm m}^{-2}~{\rm s}^{-1}~{\rm deg}^{-2}$ comes from sources that are below the X-ray flux limit.

The diffuse X-ray background contribution from lobes of powerful double radio sources is found to make up $0.02$ of the unresolved CXB. IC scattering of the CMB off the nucleus and hotspots of sources as well as X-ray clusters and FR I objects also contribute to the CXB.

\section{Volume filling factor of the universe by lobes}\label{sec:vff}

Next the volume filling factor $\zeta$ of all lobes and relic lobes is calculated at each redshift.
\begin{equation}
\zeta=\displaystyle\int_0^T 2 V_{\rm l}(t) p_{\rm A}(t,z)\,dt \times
(1+z)^3\rho_{\rm tot}(z)  
\end{equation}
gives the volume filling factor at $z$ for sources with age less than $T$ (again, the limits of integration altered appropriately determine which objects we care about). 
$T$ is set to the age of the universe at redshift $z$ if we care about all objects, and is altered appropriately to account for just active lobes or IC ghosts. We allow for expansion of the volume $V_{\rm l}(t)$ only until the time the lobe pressure comes into equilibrium with the ambient pressure, calculated from the ambient density profile assuming a temperature of $10^5~{\rm K}$
(temperature of the warm-hot intergalactic medium \citep{2006ApJ...650..560C}), at which point the volume remains constant. The ratio of the lobe pressure at time $t$ to the ambient pressure at the distance of the end of the lobe at time $t$ stays well above $1$ for times beyond which the lobe has Mpc scale length.
The lobe volume and source density will depend on jet power $Q_{\rm j}$, so in order to find $\zeta$ we bin jet power, calculate $\zeta$ at each and take a weighted average according to the empirically-inferred distribution of jet powers.
The factor of $(1+z)^3$ in the expression for $\zeta$ comes from converting comoving volume to proper volume.

\begin{figure}
\centering
\includegraphics[width=0.47\textwidth]{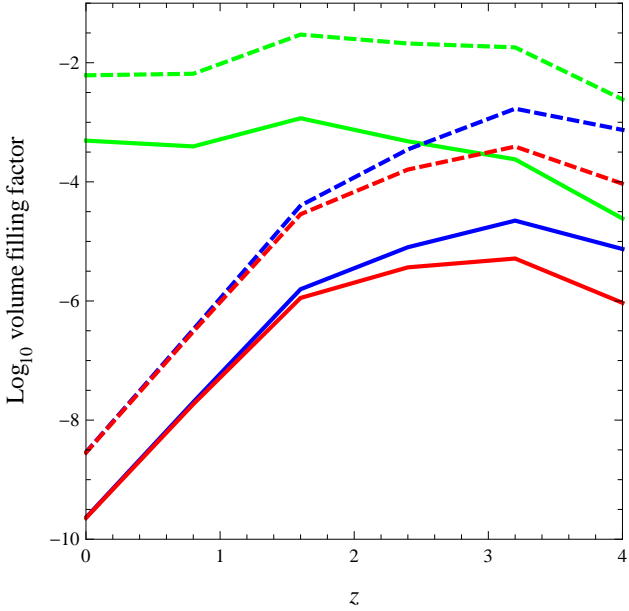}
\caption{Volume filling factors of sources as a function of redshift.
The solid lines from top to bottom refer to:
[1/green] all radio galaxies including completely X-ray and radio dark ones,
[2/blue] IC ghosts and sources with radio jets on,
[3/red] IC ghosts. 
The corresponding dashed lines show the volume filling factors if we consider the relevant volume of the filamentary structures of the universe.}
\label{fig:vff}
\end{figure}

We also consider the relevant volume the lobes are situated in, namely the volume of filamentary structures of the universe.
We divide the volume filling factor by lobes by the volume fraction of the warm-hot intergalactic medium in figure~1 of \cite{2006ApJ...650..560C} to account for the relevant volume.
Figure~\ref{fig:vff} shows the volume filling factors as a function of $z$.  
At $z=2$ and $z=3$, we predict volume filling factors of all lobes around $0.03$ and $0.02$ respectively.
For a longer jet lifetime, $t_{\rm j}=5\times10^8~{\rm yr}$, the predicted volume filling factors at these redshifts are $0.3$ and $0.2$.
While lobes may not significantly fill the entire universe, it is possible that they fill a large fraction of the filaments in which they are located.

\cite{2001ApJ...560L.115G} give a simple estimate for the volume filling factor by lobes as well. 
Their paper estimates $\zeta=0.01$ for $t_{\rm j}=10^8~{\rm yr}$ and  $\zeta=0.53$ for $t_{\rm j}=5\times10^8~{\rm yr}$ (at $z=2$).
\cite{2001ApJ...560L.115G} do not use a birth function of radio sources and estimate the number of dead lobes from $t_{\rm qe}/t_{\rm j}$, where $t_{\rm qe}$ is the duration of the quasar era. This estimate gives slightly higher ratios at $z=2$ than estimated by the birth function due to the simpler age distribution of galaxies.
Typical lobe volumes are larger in \cite{2001ApJ...560L.115G} than estimated by our model since their paper assumes a constant axial ratio of $2.5$ (according to this paper's definition of axial ratio), i.e. assuming more spherical sources.
The question of volume filling factor of lobes is also addressed with more care in the 
works by \cite{2007ApJ...658..217B} and \cite{2008ApJ...682L..17B}, with which the results of our work agree.
The former paper investigates the volume filling factor by lobes using the models of \cite{1997MNRAS.286..215K},  \cite{1999AJ....117..677B} and \cite{2002A&A...391..127M}, and also modifications by incorporating a variable hot spot
size growing with the source age. Table~7 of their paper 
gives the relevant volume fraction results for the models with a number of parameters varied. A wide range of relevant
volume filling factors can be produced by modification of the parameters, with the cumulative relevant volume filling factor of radio galaxies over the quasar era around $0.05$.
The later paper incorporates radio lobe growth into a numerical cosmological evolution, and finds a volume filling factor of $0.10$--$0.30$.

The total number of particles placed into the IGM because of the formation of the lobes is
\begin{equation}
N_{\rm particles} = \displaystyle\int_0^{t_{\rm j}} \left( d V(t_{\rm i})\displaystyle\int_{\gamma_{\rm min}}^{\gamma_{\rm max}}   n_0(t_{\rm i}) \gamma^{-p}  \,d\gamma  \right) \,d t_{\rm i}.
\label{eq:numPart}
\end{equation}
$N$ is found to be between $10^{65}$ to $10^{67}$ particles for the different jet energies. The density of these particles in the lobes at the time the jet turns off is on the order of $10^{-32}$--$10^{-31}~{\rm kg}~{\rm m}^{-3}$. 
Therefore, the ratio of the density of particles in the lobes to the critical density of the universe $\Omega_{\rm radio~lobes}$ is at least $6$ orders of magnitude below unity, even for a lobe volume filling factor of $1$. 

\section{X-ray luminosity function of extended X-ray sources}\label{sec:XLF}

\begin{figure}
\centering
\includegraphics[width=0.47\textwidth]{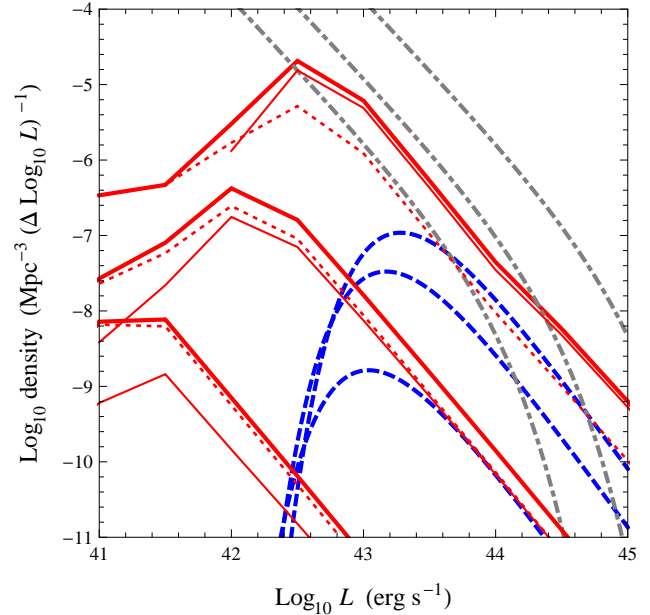}
\caption{The radio ($151~{\rm MHz}$) and X-ray ($1~{\rm keV}$) luminosity function for various sources. The bottom, middle, top dashed (blue) lines refer to the RLF of FR II objects derived from the 7CRS, 6CE and 3CRR surveys at $z=0.2,1,2$ respectively. The bottom, middle, top thick solid (red) lines refer to the XLF of all extended X-ray sources derived from our model $(t_{\rm j}=10^8~{\rm yr})$ and age and density distribution at $z=0.2,1,2$ respectively. Dotted thin lines show the IC ghost part of the population and solid thin lines shows the currently active galaxy population. For comparison, the XLF of X-ray clusters \citep{2004ApJ...607..175M} (dot-dashed/grey) is shown (top, middle, bottom correspond to $z=0.2,1,2$ respectively).}
\label{fig:rlf}
\end{figure}

The X-ray luminosity function for extended X-ray sources may be directly found from our model since we have described the density and the age and jet energy distributions of the sources and have a model for the evolution of the luminosity. We are not required to take a time-average of ratios as in \cite{2010MNRAS.tmp.1004N} to convert the RLF to an XLF. We present in Figure~\ref{fig:rlf} the XLF for all extended X-ray sources derived from our model and synthetic population. For comparison, the XLF for X-ray clusters from \cite{2004ApJ...607..175M} is also plotted. The evolution of the cluster XLF is not well known at luminosities $< {\rm few} \times 10^{44}~{\rm erg}~{\rm s}^{-1}$ and beyond redshift $z=0.8$, however, observed volume densities for $0.6<z<0.8$ and for luminosities above ${\rm few} \times 10^{44}~{\rm erg}~{\rm s}^{-1}$ are significantly lower than those at the present population \citep{2004ApJ...607..175M}.

The XLF derived from our model, when integrated, gives a higher number density than does integrating the original RLF because it accounts for the predicted total number of luminous sources, whereas, we have mentioned earlier, the RLF has no knowledge of the evolution of radio sources and therefore when integrated only accounts for the density of the sources above the lowest flux limit of the samples used in its construction.

\section{Varying model parameters}\label{sec:params}

\begin{figure}
\centering
\includegraphics[width=0.47\textwidth]{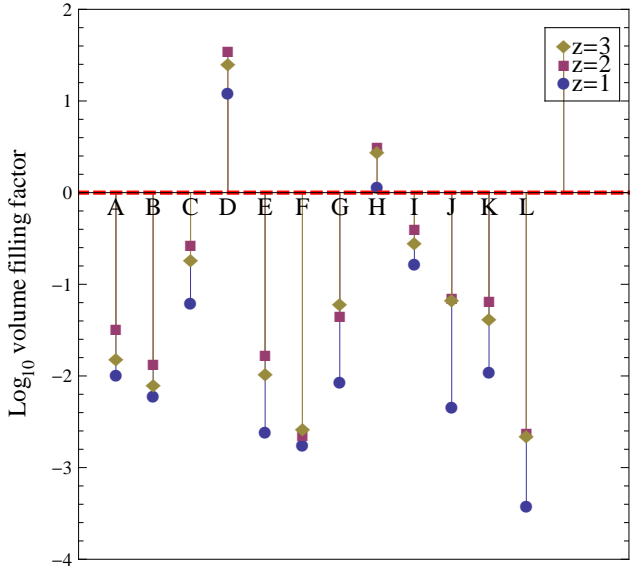}
\caption{Volume filling factors of the filamentary structure of the universe by lobes at redshifts $z=1,2,3$ for various model parameters. Case
[A] original parameters investigated ($t_{\rm j}=10^8~{\rm yr}$, $p=2.14$, $\gamma_{\rm min}=1$, $\rho_0= 1.67\times 10^{-23}~{\rm kg}~{\rm m}^{-3}$, $\beta=1.5$),
[B] $t_{\rm j}\to 5\times10^7~{\rm yr}$,
[C] $t_{\rm j}\to 5\times10^8~{\rm yr}$,
[D] $p\to 3$,
[E] $\gamma_{\rm min}\to 2000$,
[F] $p\to 3$ and $\gamma_{\rm min}\to 2000$,
[G] $p$ correlated with $Q_{\rm j}$,
[H] $\rho_0\to 1.67\times 10^{-24}~{\rm kg}~{\rm m}^{-3}$,
[I] $\beta\to 2$,
[J] birth function $\Delta z\to 1$,
[K] jet-power power-law $-2.6\to -3.0$,
[L] jet-power power-law $-2.6\to -2.0$.
The horizontal dashed line shows a volume filling factor of unity. In some instances the predicted volume filling factor exceeded unity, which means that the ejected material from different sources overlap.
}
\label{fig:bgvff}
\end{figure}

\begin{figure}
\centering
\includegraphics[width=0.47\textwidth]{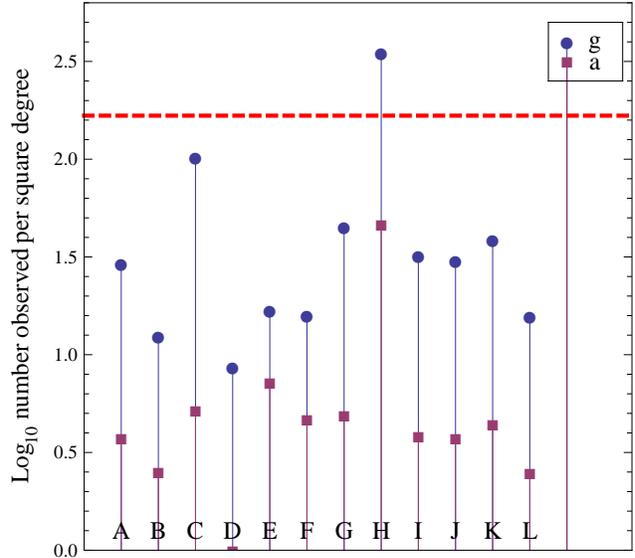}
\caption{Number of sources predicted to be visible above X-ray flux limit of CDFN survey for various model parameters. `a' stands for all sources (switched on and switched off), `g' stands for IC ghosts.
[A]--[L] same as in Figure~\ref{fig:bgvff}.
The horizontal dashed line shows the number of extended X-ray sources expected to be observed above X-ray flux limit of CDFN survey.
}
\label{fig:bgno}
\end{figure}

\begin{figure}
\centering
\includegraphics[width=0.47\textwidth]{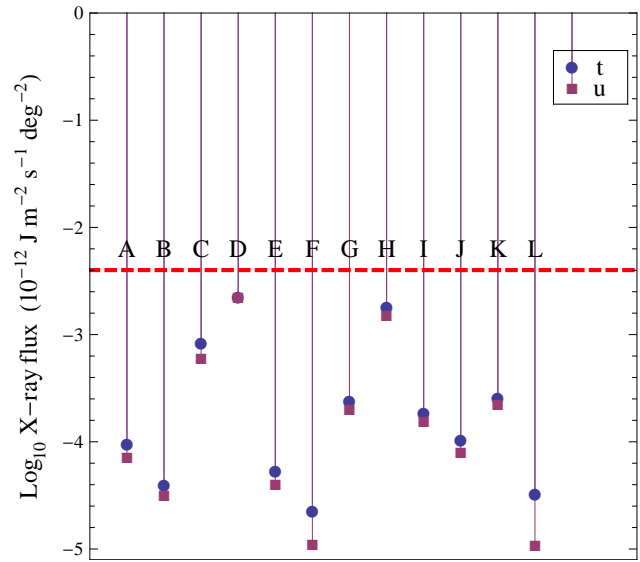}
\caption{Expected X-ray flux from radio sources expected for various model parameters. `t' stands for the total flux, `u' stands for the unresolved part (i.e. due to objects below the X-ray flux limit).
[A]--[L] same as in Figure~\ref{fig:bgvff}.
The horizontal dashed line shows X-ray flux from the CXB after all resolved X-ray point and extended sources have been removed.
}
\label{fig:bgxf}
\end{figure}

We test how sensitive our predictions are to changes in the model parameters.
Varying the parameters can indeed have a significant effect on the distribution of double-lobed sources and IC ghosts, since different model parameters change the times sources go below the radio flux limit, and density parameters and jet lifetimes also increase or decrease the volumes of the sources.
The parameters we alter are the jet lifetime $t_{\rm j}$, the injection index $p$, the minimum energy of injected electrons $\gamma_{\rm min}$ and parameters describing the surrounding AGN environment. In addition, we also investigate the effect of the birth function.
We observe how our results change with shorter and longer jet lifetimes of $t_{\rm j}=5\times10^7~{\rm yr}$ and $t_{\rm j}=5\times10^8~{\rm yr}$ (cases [B] and [C]). 
The injection index $p=2.14$ we considered is not significantly steep so we also consider the steep case ([D]) of $p=3$ (observational work by \cite{1987MNRAS.225....1A} find $2\leq p \leq 3$). Additionally, we consider correlating $p$ with jet power $Q_{\rm j}$ (case [G]), inspired by the $P$--$\alpha$ correlation  of \cite{1999AJ....117..677B}.
A simple linear correlation between $p$ and $\log_{10} Q_{\rm j}$ is assumed, with $p=2$ corresponding to $Q_{\rm min}$ and $p=3$ to $Q_{\rm max}$.
We consider injected electrons to have $\gamma_{\rm min}=2000$ (cases [E] and [F]), greater than the $\gamma\sim10^3$ factors required for the X-ray radiation, in addition to the original case ([A]) which had $\gamma_{\rm min}=1$.
We also examine the importance of the surrounding density and the effect of reducing the density by an order of magnitude by setting $\rho_0=1.67\times 10^{-24}~{\rm kg}~{\rm m}^{-3}$ (case [H]), as well as considering an ambient density profile given by a steeper power law with $\beta=2$ (case [I]).
We consider a wider birth function with $\Delta z=1$ (case [J]).
Finally, we consider modifying the power-law distribution of jet energies to $-3.0$ and $-2.0$ (cases [K] and [L]).
In Figures~\ref{fig:bgvff},~\ref{fig:bgno},~\ref{fig:bgxf} we present how the volume filling factor, source number density above the X-ray flux limit and the unresolved X-ray emission change with varying the model parameters.

In the original case [A] we predict $\sim29$ lobed-galaxies per square degree visible in the X-ray, of which $\sim4$ are IC ghosts, above the X-ray flux limit of the CDFN survey. The radio sources contribute about $2$~per~cent of unresolved CXB and the volume filling factor of the lobes is $\sim0.03$--$0.02$ at $z=2$--$3$.

The jet lifetime is an important parameter that affects the predicted distribution of the sources. A larger jet lifetime means that sources will grow longer while there are also fewer predicted sources with jets turned off during the quasar era.
The predicted volume filling factor exceeds $0.25$ at redshifts of the quasar era for an increased jet lifetime of $5\times 10^8~{\rm yr}$, while the volume filling factor decreases to $\zeta\leq 0.01$ for a decreased jet life time of $5\times 10^7~{\rm yr}$. 
Evidence that a significant fraction of sources have long jet lifetimes on the order of $5\times 10^8~{\rm yr}$ will mean that double-lobed sources can account for $20$~per~cent of the unresolved CXB and $60$~per~cent of the extended sources in the CDFN survey.

Increasing the injection index to $p=3$ has the effect of reducing the radio luminosities and consequently sources are more likely to be below the radio flux limit and hence a larger number of total sources in the underlying distribution are predicted. Consequently, the predicted volume filling factor of sources at redshift $z$ are significantly higher, surpassing unity in the quasar era and even at $z=1$. The contribution of lobes to the unresolved CXB is $60$~per~cent (Figure~\ref{fig:bgxf}).
Despite the large volume fraction, the sources are less bright in the X-ray compared with [A] and the number of sources observed per square degree is $9$ at the flux density limit of the CDFN survey.
  
The minimum electron injection energy determined by $\gamma_{\rm min}$, when increased to $\gamma=2000$, leads to higher X-ray and radio luminosities than our original case, meaning there are fewer sources predicted to go undetected and hence a smaller number of total sources in the underlying distribution and volume filling factor. The volume filling factor is still as high as $0.017$ during the quasar era.
Additionally, setting $\gamma_{\rm min}=2000$ results in the X-ray luminosity falling less steeply for the majority of the time the source is an IC ghost, meaning that it is above the flux limit for a longer period of time and hence more observable. The ratio of observable IC ghosts to all observable sources for $\gamma_{\rm min}=2000$, at the flux limit of the CDFN survey is $30$~per~cent versus $13$~per~cent for $\gamma_{\rm min}=1$. 

\begin{figure}
\centering
\includegraphics[width=0.47\textwidth]{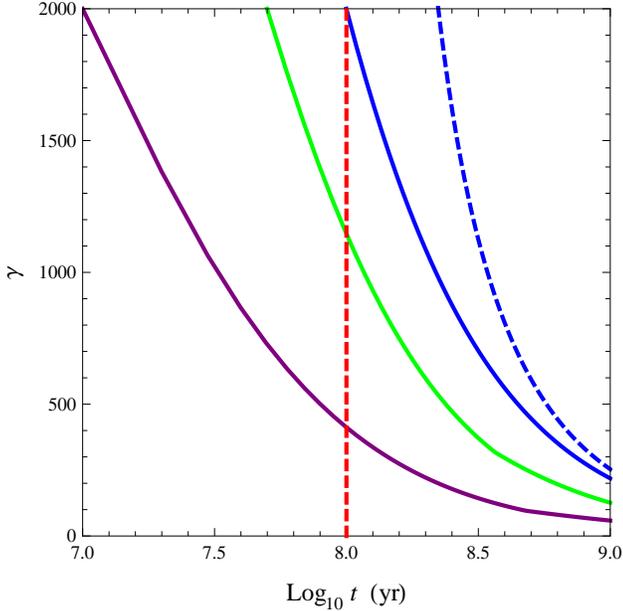}
\caption{The evolution of Lorentz factors since time of injection into the lobe.
For a jet lifetime of $t_{\rm j}=10^8~{\rm yr}$, $Q_{\rm j}=10^{38}~{\rm W}$, $z=2$ we show how a particular $\gamma$ evolves as a function of time. We plot the evolution of $\gamma_{\rm min}=2000$ when it is injected into the lobe at times $0.1 t_{\rm j}$ (purple), $0.5 t_{\rm j}$ (green) and $t_{\rm j}$ (blue). We also show the evolution of $\gamma_{\rm max}=10^6$ injected at time  $t_{\rm j}$ (blue, dashed), defining an upper limit of observable $\gamma$ factors after the jet stops.
The vertical dashed line shows the time the jet switches off.
}
\label{fig:gEvo}
\end{figure}

By changing both $p$ and $\gamma_{\rm min}$ appropriately it may be possible to have a typically high injection index and predict reasonable values for the observable distribution (number density, CXB contribution, volume filling factor, especially after the quasar era) of double-lobed sources. Such is the case when $p=3$ and $\gamma_{\rm min}=2000$, where we are not underestimating radio luminosities as in the $p=3$ and $\gamma_{\rm min}=1$ case. The minimum injected Lorentz factor $\gamma_{\rm min}$, whose typical value is not well known, appears to be an important parameter for understanding the luminosity evolution of powerful double-lobed sources. Figure~\ref{fig:gEvo} shows the evolution of Lorentz factors since the time of injection.

In correlating injection index $p$ with jet power $Q_{\rm j}$, we have made the least powerful sources brighter and the most powerful sources dimmer. This could, in principle, alter the volume filling factor and number density in either direction. For our correlation we described earlier in this section, the number density and volume filling factor are increased somewhat from the original case, and double-lobed sources account for $27$~per~cent of the extended sources of the CDFN survey and $6$~per~cent of the unresolved CXB.

The ambient density of galaxies and its evolution, which is still not well understood, could significantly affect the estimate for the volume filling factor of filaments by radio lobes. Decreasing the ambient density (reducing $\rho_0$ by a factor of $10$) allows for sources to grow larger and initially more luminous, but luminosity also falls more quickly due to increased adiabatic losses. Volume filling factors are increased from the original case to above unity. The observable number density of sources is higher by a factor of $2$, however, than observed by the CDFN survey. The unresolved contribution to the CXB is $45$~per~cent.
We also investigate the effect of a steeper density profile with $\beta=2$. The surrounding density is larger than the original case for distances $r<a_0=10~{\rm kpc}$ and smaller at distances $r>a_0$. The lobes will end up growing much larger and radio luminosity falls much more quickly in the evolution of the source. Volume filling factors are increased from case~[A] to $\sim0.4$. The contribution to the unresolved CXB is $5$~per~cent.
It may be the case that the typical density profile around FR II sources falls more steeply at higher radii, such as if the density profile is similar to that of relaxed clusters \citep{2006ApJ...640..691V}, which follow the Navarro-Frenk-White halo profile (less steep than $r^{-2}$ near the centre, more steep than $r^{-2}$ at large distances) \citep{1996ApJ...462..563N, 1997ApJ...490..493N}.
Since most sources at redshifts $z\geq 2$ tend to fall below the radio flux limit early in their evolution before they reach Mpc scales (their fractional duration of visibility compared to the jet life time may be less less than $0.01$-$0.1$) \citep{1999AJ....117..677B}, and the total density of double-lobed sources depends on how quickly the sources fall below the flux limit, the density profile nearer the source rather than at large distances plays a more important role in determining the total population of double-lobed sources.

If most sources were born over a broader span of time, which we investigate by setting $\Delta z=1$, the volume filling factor during the quasar era is slightly larger than estimated with our original parameter. This is a result of the fraction of sources that would be observable above the radio flux limit decreasing due to a wider spread in ages of sources at each redshift. With $\Delta z=1$, we find a volume filling factor of $0.07$ at redshifts $2$ and $3$, assuming all the parameters are the same as in our original case [A], while the number density and unresolved CXB contribution are comparable to the population described by the original birth function with $\Delta z=0.6$.

Steepening the power-law index of jet energies from $-2.6$ to $-3$ has the effect of a larger number of sources being undetected in the survey and flattening it has the opposite effect. 
Compared to the original case [A] which predicts $\sim29$ lobed-galaxies per square degree visible in the X-ray, of which $\sim4$ are IC ghosts, above the X-ray flux limit of the CDFN survey, and predicts the radio sources contribute about $2$~per~cent of unresolved CXB and the volume filling factor of the lobes is $\sim0.03$--$0.02$ at $z=2$--$3$,
if we increase the power-law index of jet energies from $-2.6$ to $-3$ we find
$\sim 39$ visible sources per square degree of which $\sim4$ are IC ghosts, a $6$~per~cent contribution to the unresolved CXB, and a volume filling factor of the lobes is $\sim0.07$--$0.04$ at $z=2$--$3$.
And if we decrease the power-law index of the jet energies to $-2$, 
we find
$\sim 16$ visible sources per square degree of which $\sim2$ are IC ghosts, a $1$~per~cent contribution to the unresolved CXB, and a volume filling factor of the lobes is $\sim0.01$-$0.004$at $z=2$--$3$.


\section{Discussion and Conclusion}\label{sec:Discussion}
We have developed an analytic model to study the evolution of the radio luminosity and X-ray luminosity (due to IC scattering of the CMB) of FR II radio galaxies in order to quantify the abundance of actual and observable powerful double-lobed radio sources and IC ghosts (sources with jets turned off that still radiate in the X-ray).
For a set of model parameters inferred from observations of radio sources (case [A]), which had 
$t_{\rm j}=10^8~{\rm yr}$, $\gamma_{\rm min}=1$, $p=2.14$, $\rho_0= 1.67\times 10^{-23}~{\rm kg}~{\rm m}^{-3}$ and $\beta=1.5$,
we predict $\sim29$ lobed-galaxies per square degree visible in the X-ray, of which $\sim4$ are IC ghosts, above the X-ray flux limit of the CDFN survey. 
The CDFN survey found $167^{+97}_{-67}$ extended X-ray sources, which consist of both X-ray clusters and lobes. Thus a considerable fraction of these objects may turn out to be in fact powerful radio lobes based on our estimate.
It is important to note that the region observed with the CDFN survey may not be a typical area of the X-ray sky because the survey was chosen to follow up on the Hubble Deep Field North observations, which looked at an area of the sky free from bright emissions in the visible, radio, infrared, ultraviolet and X-ray.
It is also expected by our model that the powerful radio sources will contribute about $2$~per~cent of unresolved CXB, which is due to IC scattering of the CMB by the lobes and nuclei of individual galaxies as well as clusters.
The volume filling factor of the lobes is highest during the quasar era, at $\zeta\sim0.03$ for $z=2$. However, we also find that our predictions for volume filling factor are sensitive to the model parameters we use. Especially important could be the minimum injection energy, determined by $\gamma_{\rm min}$, whose typical value is not well known. In addition, a higher value of $\gamma_{\rm min}$ also improves the visibility of IC ghosts and can increase the observable ratio of IC ghosts to active sources by a factor of $3$.

For case [A] the volume filling factor is $\lesssim 0.03$, highest at $z\sim2$. The volume filling factor at the quasar era is about an order of magnitude above that of the present era. The volume filling factor declines only slightly from $z=2$ to $z=3$.
The volume filling factor of dead lobes is much larger than that of active ones by at least an order of magnitude at all redshifts, but especially at low redshifts $z<1.5$.

The derived XLF at $1~{\rm keV}$ of double-lobed sources (both active and ghosts) shows that they are more abundant at higher redshifts ($z\gtrsim 1$) than at lower redshift and are also generally more luminous in the X-ray at these higher redshifts (the peak of the XLF shifts to the right as redshift is increased). Additionally, at redshifts $z\gtrsim 2$ the space densities of double-lobed radio galaxies become comparable to that of X-ray clusters at high luminosities ($L_{\rm x}\geq {\rm few}\times 10^{43}~{\rm erg}~{\rm s}^{-1}$) - see Figure~\ref{fig:rlf}. This is even true of just the high redshift IC ghosts at high luminosities ($L_{\rm x}\geq {\rm few}\times 10^{44}~{\rm erg}~{\rm s}^{-1}$).

In this paper we have only considered FR II sources and have neglected FR I sources, which are much more numerous in the local universe than FR IIs. \cite{2008MNRAS.388.1335W} creates a semi-empirical simulation of the extragalactic radio continuum sky
out to redshift $z=20$, and down to flux density limits of $10$~nJy, which includes FR I and FR II simulated sources drawn from the RLF of \cite{2001MNRAS.322..536W} and shows that across all redshifts, FR I sources can be $5$ orders of magnitude more abundant that FR II sources.
However, the RLF of \cite{2001MNRAS.322..536W} shows that FR I and FR II sources have comparable number density 
during the redshift range ($z=2-3$), the quasar era, on which we are focusing. 
Accessing simulation data products of \cite{2008MNRAS.388.1335W} through the
data base web interface (http://s-cubed.physics.ox.ac.uk) also agrees with there being a comparable number density during the quasar era.
The relative contribution from FR I sources during the quasar era, for example to the volume filling factor, is modest. 
The volume of a typical 
FR I will be much smaller than the volume of an FR II source, since
FR I seem to have a much higher efficiency in converting jet thrust to radio flux and 
are also much dimmer (see
\cite{2007ApJ...658..217B} and references therein).
In assessing the importance of FR I objects to the CXB, the question at hand is the integrated number of $\gamma\simeq 10^3$ particles that the FR I produces during its lifetime.
The number of particles ejected by an FR I will be less than that of an FR II object
since the FR II sources have higher luminosity and jet power, and lower efficiency in converting beam power to radio flux.
In addition, FR Is may have higher typical B-fields, which would reduce the number of particles in the lobe needed to produce the object's observed luminosity.  
The environments of an FR I object may be more magnetised. See, for example \citet{2008MNRAS.391..521L} and
\citet{2008ASPC..386..104L} where the FR I object 3C 31's magnetoionic environment is thought to have a B-field of $0.21$~nT.
This value is a lower limit to the B-field that 3C31's jet plasma is bathed in. The actual B-field in the lobes is likely higher since FR~I's are typically very lossy and local shear effects will only heighten the field strengths.
While there are individual examples of FR Is that are exquisitely well-modelled in terms of velocity profile, etc. (see for example, \cite{2008MNRAS.386..657L}, \cite{2008MNRAS.391..521L}) the modelling of typical FR Is to date is far behind that of FR IIs. In a future work, it would be worthwhile to develop a model of FR I sources with comparable physical detail as the models for FR II sources to assess their contribution to the unresolved CXB.

It is possible that typical double-lobed sources have multiple outbursts of jet activity, while our model only accounted for a single outburst. If sources show intermittent activity as in \cite{1997ApJ...487L.135R}, active sources having fewer past outbursts will be brighter than those with a long history of episodes of jet ejection and hence more easily detectable. If during the jet lifetimes we assumed the source switches off a number of times then the lobes grow smaller in volume. Since sources tend to fall below the radio flux limit during the quasar era early in their lifetimes, the predicted total density of double radio sources will not be altered significantly by the inclusion of multiple outbursts (unless periods of jet activity are shorter than time at which the source falls below the radio flux limit). Therefore, the volume filling factor will be somewhat reduced. If, on the other hand, sources have recurring extended periods of jet activity of the order of the jet lifetimes we assumed, then lobes will grow larger still and the volume filling factor would increase.

Our predicted total comoving density of active and non-active double radio sources at $z=1$ and $2$ is similar to the comoving density of galaxies above a luminosity $M_{\rm K}=-25$ and $-24.5$ in the galaxy luminosity function of \cite{2010MNRAS.401.1166C}. The predicted total comoving density of FR II sources suggests that most massive galaxies in the universe may have an FR II type outburst at some point in their evolution.
Using the correlation between K-band bulge luminosity and black hole mass for galaxies given by \cite{2003ApJ...589L..21M}, 
the predicted total comoving density of active and non-active double radio sources is similar to the comoving density of all the brightest galaxies with supermassive black holes of masses $\geq 10^{8.7}~{\rm M}_\odot$ and $10^{8.4}~{\rm M}_\odot$. \cite{2002MNRAS.331..795M} argue that the true radio loud AGN have black hole masses $\geq 10^{8.5}~{\rm M}_\odot$. All these galaxies may at some point in their history have contributed to returning material into the IGM in large-scale outbursts.

We have studied the actual and observable distribution of active double-lobed galaxies, IC ghosts and dead lobes. IC ghosts may comprise $10$--$30$~per~cent of double-lobed X-ray sources and could outnumber X-ray clusters at high luminosities. 

The volume filling factor of all lobes calculated from the wide range of model parameters investigated suggests that it is likely between $0.001$--$1$ during the quasar era, and the contribution to the unresolved CXB is $1$--$60$~per~cent. 
It is important to know the parameters of a typical radio source well to be able to constrain the inferred distribution,  as we see that our predictions have some extreme sensitivities to
underlying physical parameters  (\S~\ref{sec:params}).
Having a significant fraction of the relevant volume of the universe permeated by overpressured lobes of galaxies can trigger large scale star formation and seed filaments with magnetic fields, as summarized by \cite{2001ApJ...560L.115G}. The volume filling factor at $z=3$ is comparable to the volume filling factor at $z=2$, meaning that radio lobes may have played an important role in star formation from an early time for a long-range period. Parameters of individual sources, such as $\gamma_{\rm min}$, may significantly influence the cosmological predictions for volume filling factor and radiation contribution to the unresolved CXB, and therefore it is important to obtain better constraints on these parameters from observations.

\section*{Acknowledgments}
PM would like to acknowledge the award of a Weissman grant from Harvard University. KMB and ACF thank the Royal Society for support. This research made use of the NASA/IPAC Extragalactic Database which is operated by the Jet Propulsion Laboratory, Caltech, under contract with the National Aeronautics and Space Administration.

\bibliography{mybib}{}

\bsp

\label{lastpage}

\end{document}